\documentclass[prb,longbibliography,twocolumn]{revtex4-1}

\usepackage{graphicx}
\usepackage{dcolumn}
\usepackage{bm}
\usepackage{hyperref}
\usepackage{amsmath, amssymb}
\usepackage{mathtools}
\usepackage{color}
\usepackage[none]{hyphenat}

\hypersetup{pdftex,
  breaklinks=true,
  colorlinks=true,
  urlcolor=blue,
  linkcolor=blue,
  citecolor=blue,
  }

\begin{document}


\title{Metal-Coordination Effects on the Stability and ORR/OER Activity of Layered Organometallic Single-Atom Catalysts: A Theoretical Study}

\author{Pedro H. Souza}
\author{Victor Hoyos-Sinchi} 
\author{Walter Orellana}
\email{worellana@unab.cl}

\affiliation{Departamento de F\'isica y Astronom\'ia, Facultad de Ciencias Exactas, Universidad Andres Bello, 
Sazi\'e 2212, Santiago 8370136, Chile}
        
\date{\today}

\begin{abstract}
Organometallic layered materials have emerged as promising single-atom catalysts for oxygen reduction and evolution reactions, 
but their practical use has been limited by insufficient electrochemical stability. Here, we present a density functional theory study clarifying 
the relationship between catalytic activity and stability in organometallic single-atom catalysts with metal-N$_4$ (MN$_4$) and 
metal-O$_4$ (MO$_4$) coordination. 
We compare graphene-embedded MN$_4$ motif and phthalocyanine-like frameworks with MO$_4$-coordination frameworks, including 
M$_4$(OHPTP)$_2$ and M$_3$(HHTP)$_2$ (M = Mn, Fe, Co, Ni, Cu, Zn). Stability is assessed by surface Pourbaix analysis, while 
activity is evaluated using the computational hydrogen electrode method. MN$_4$ systems show competitive overpotentials but suffer 
strong pH-dependent instability. In contrast, MO$_4$ frameworks exhibit enhanced robustness across wide pH ranges while maintaining 
good catalytic performance. A proposed stability descriptor enables direct comparison across systems, identifying MO$_4$ coordination 
structures, particularly M$_4$(OHPTP)$_2$ (M = Zn, Co) as optimal for balancing activity and stability in practical electrocatalysis.
\end{abstract}

\maketitle

\section{Introduction}
The development of sustainable energy conversion technologies critically depends on efficient electrochemical processes, 
particularly the oxygen reduction reaction (ORR) and oxygen evolution reaction (OER). A central challenge in this field is the 
replacement of noble-metal-based catalysts, which exhibit excellent activity but face significant limitations for large-scale 
applications due to their high cost and limited natural abundance.

Metal cations coordinated by pyridinic nitrogen sites embedded in graphene (G-MN$_4$) have been widely reported as 
single-atom catalysts for both ORR and OER.\cite{jin2021,jli2021,cwan2020} However, 
their catalytic performance is strongly metal dependent and highly sensitive to the electrolyte environment. For instance, 
G-MnN$_4$ exhibits favorable ORR activity in alkaline media but shows markedly reduced performance under acidic 
conditions.\cite{li2024, zhu2019} Similarly, G-FeN$_4$ display substantial variations in both activity and stability as a 
function of pH, driven by changes in reaction intermediates and local electronic structure.\cite{meng2009, yang2021} 
G-CoN$_4$  and G-NiN$_4$ structures also demonstrate catalytic activity; however, their operation is typically confined 
to narrow pH windows.\cite{gong2023} Collectively, these observations indicate that, despite their intrinsic activity, 
G-MN$_4$ structures suffer from pronounced pH-dependent limitations that significantly restrict their practical applicability 
across diverse electrolyte environments.\cite{kumar2026}

On the other hand, two-dimensional metal-organic frameworks (2D MOFs) have emerged as promising materials for 
electrocatalysis owing to their well-defined coordination environments, high density of active sites, and pronounced 
structural tunability.\cite{chan2025bi,fu2025fe,wei2022no} Within this class, 
phthalocyanine-like MOFs (s-MPc) have been investigated as model systems for ORR and OER.\cite{lv2023} 
These materials feature a well-defined MN$_4$ coordination motif and have demonstrated favorable intrinsic catalytic 
activity. Nevertheless, similar to their graphene-supported counterparts, they often exhibit limited structural stability 
under acidic and alkaline electrolyte conditions.

Another important family of 2D MOFs is represented by the general formula M$_3$(ligand)$_2$. These layered frameworks 
predominantly adopt metal bis(dioxolene) (MO$_4$) and metal bis(diamine) (MN$_4$) coordination motifs, which give rise 
to distinct electronic structures and catalytic properties. Early experimental studies demonstrated promising electrocatalytic 
performance in alkaline media, exemplified by 
Ni$_3$(HITP)$_2$ (HITP = 2,3,6,7,10,11-hexaiminotriphenylene) for the oxygen reduction reaction (ORR)\cite{miner2016} and 
Co$_3$(HHTP)$_2$ (HHTP = 2,3,6,7,10,11- hexahydroxytriphenylene) for the oxygen evolution reaction (OER)\cite{mingdao2018}. 
Notably, these materials exhibit activities comparable to those of state-of-the-art precious-metal catalysts. Subsequent studies 
have further expanded their electrocatalytic scope,\cite{feng2018,hongxing2018,jiawenli2021} encompassing 
ORR,\cite{park2020,qianzhao2021} 
OER,\cite{xiao2022,wang2024} hydrogen evolution reaction 
(HER),\cite{pathak2024,xing2025,chen2021cofe,chen2021co} as 
well as the electroreduction of CO$_2$\cite{yang2022,yi2021} and N$_2$.\cite{cui2019,li2020}

Theoretical investigations have explored the stability, electronic structure, and catalytic performance of 2D MOFs with diverse 
coordination environments, including MS$_4$,\cite{wang2022} 
MSe$_4$,\cite{xu2023two} 
MN$_4$,\cite{xuejingyang2018,das2023bi,orellana2025} 
MO$_4$,\cite{wang2023spin,wei2024un,hoyos-sinchi2025,souza2025} and mixed 
motifs.\cite{zhou2024me,youxiwang2025,zhou2025} Collectively, these studies identify 
the metal-ligand coordination environment as a key descriptor governing electronic structure, adsorption energetics, and 
catalytic activity. Nevertheless, electrochemical stability and degradation under realistic operating conditions remain relatively 
underexplored.\cite{gorgen2022,orellana2026}

In this work, density functional theory (DFT) calculations are employed to investigate the electronic properties, electrochemical 
stability, and ORR/OER catalytic activity of single-layer organometallic materials featuring metal-N$_4$ and metal-O$_4$ 
coordination motifs. The metal-N$_4$ systems considered include graphene-embedded MN$_4$ moieties (G-MN$_4$) and 
phthalocyanine-like MOFs (s-MPc), whereas the metal-O$_4$ counterparts comprise M$_4$(OHPTP)$_2$ 
(OHPTP = 2,3,6,7,10,11,14,15-octahydroxytriphenylene) and M$_3$(HHTP)$_2$ frameworks, with M = Mn, Fe, Co, Ni, Cu, and Zn.
Our results suggest that nitrogen-coordinated systems can achieve competitive catalytic activity but are more prone to 
electrochemical degradation. In contrast, oxygen-coordinated systems combine favorable ORR/OER activity with greater 
electrochemical stability, highlighting the key role of the metal coordination environment.

\section{Computational details}

Spin-polarized density functional theory (DFT) calculations were performed using the Vienna Ab initio Simulation Package 
(VASP 6.5).\cite{vasp} Geometry optimizations and electronic structure calculations were performed using the vdW-DF2 
functional, which incorporates the GGA-type PW86 exchange functional.\cite{DF2} Self-interaction effects 
in the 3$d$ states of transition metals were treated using the Hubbard $U$ approach. The $U$ values 
assigned to  Mn, Fe, Co, Ni, Cu, and Zn were set to 3.1, 3.3, 3.3, 3.4, 3.5, and 3.5 eV, respectively, based on prior studies.
\cite{hamada2018} This treatment is critical for accurately capturing adsorbate-metal interactions, which strongly 
depend on the metal $d$-orbital character. A vacuum region of 15~\AA\ was introduced along the $z$ direction for all 
unit-cell structures. All simulations employed a plane-wave energy cutoff of 400~eV, with Brillouin zone sampling performed 
using a $2\times2\times1$ $k$-point mesh. Full geometry optimization was performed until the residual forces on all atoms 
were below 0.025~eV/\AA, with electronic self-consistency achieved at a convergence threshold of 10$^{-5}$~eV. Data 
processing and visualization were performed using the VASPKIT \cite{vaspkit} and VESTA \cite{vesta} software packages, respectively.

Electrochemical stability under electrolyte conditions was assessed through surface Pourbaix analysis. The standard 
oxidation potential of the metal sites was estimated from the surface work function. This approach enables the construction 
of potential-pH stability diagrams, providing a systematic framework to evaluate how electrode potential and electrolyte 
conditions influence surface oxidation.\cite{orellana2025} The thermodynamic overpotentials associated with the ORR and 
OER at the metal centers were evaluated using the computational hydrogen electrode (CHE) approach developed by N{\o}rskov 
and co-workers.\cite{norskov2004} Within this formalism, the adsorption free energies of the key reaction intermediates OH*, 
O*, and OOH* were calculated as $\Delta G = \Delta E_{\rm DFT} + E_{\rm ZPE} - T \Delta S$. 
where $\Delta E_{\rm DFT}$ is the adsorption energy obtained from DFT calculations, $E_{\rm ZPE}$ is the zero-point 
energy correction, and $T\Delta S$ is the vibrational entropy contribution. The latter terms were derived from vibrational 
frequency calculations of the adsorbed intermediates, considering only their vibrational degrees of freedom while keeping 
the catalyst surface fixed. Thermodynamic and vibrational parameters for the gas-phase reference molecules were taken 
from the NIST database. To account for solvent effects, adsorption free energies were corrected for hydrogen-bonding 
interactions involving the OH* and OOH* intermediates, which are stabilized through solvation. These corrections were 
evaluated for each structure using the continuum dielectric model implemented in VASPsol.\cite{mathew2014}
    
\section{Results and Discussions}
\subsection{Structural and Electronic Properties}
\begin{figure*}[t]
\includegraphics[width=17cm]{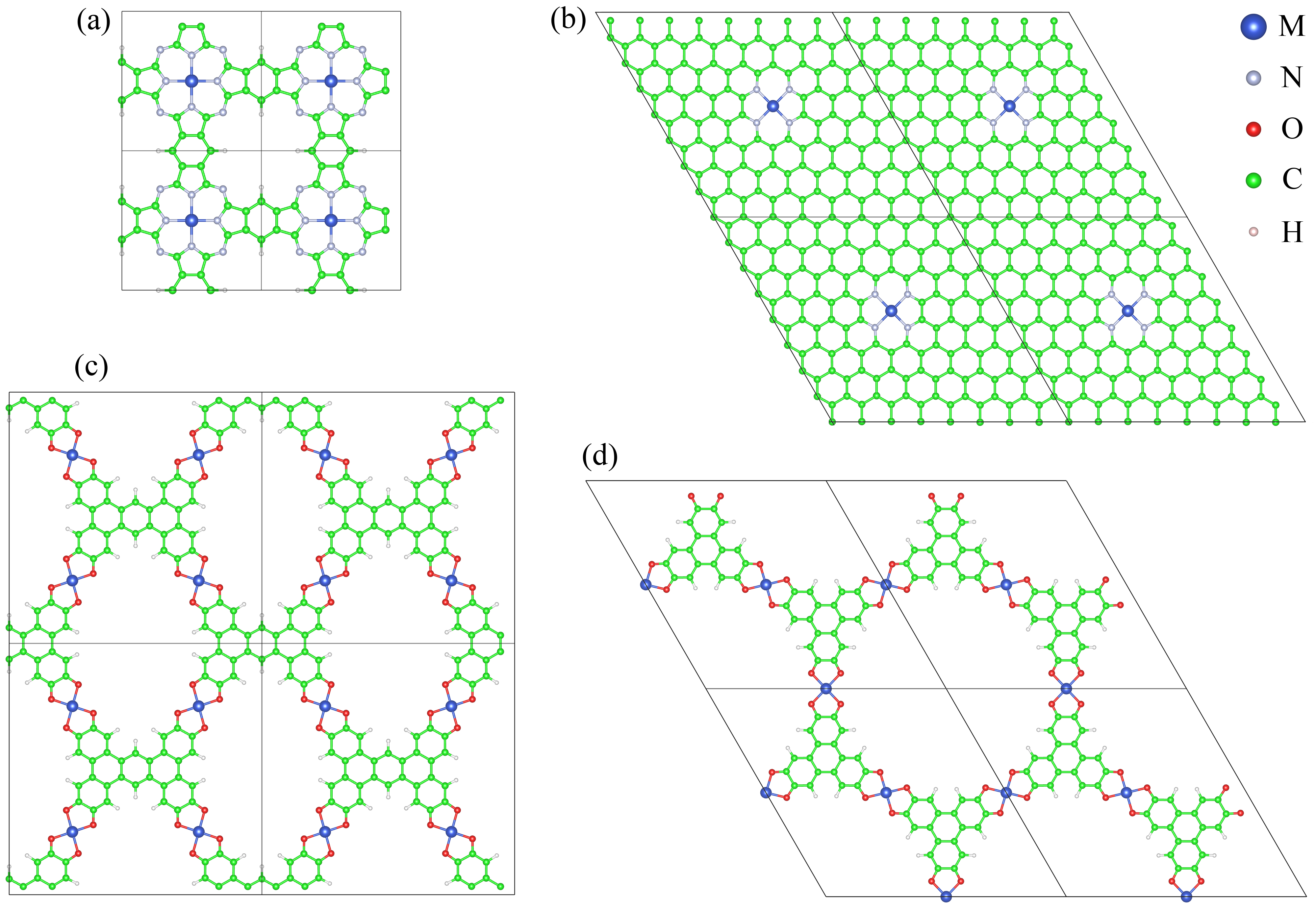}
\caption{Atomic geometry of single-layer organometallic structures in the 2$\times$2 surface unit-cell representation. 
(a) s-MPc, (b) G-MN$_4$, (c) M$_4$(OHTPT)$_2$, and (d) M$_3$(HHTP)$_2$, with M = Mn, Fe, Co, Ni, Cu, Zn. }
\label{f1}
\end{figure*}
Figures~\ref{f1} show the atomic structures of the metal-N$_4$ and metal-O$_4$ systems investigated in this work, represented 
within a 2$\times$2 surface unit cell. Figures~\ref{f1}(a) and \ref{f1}(b) correspond to the metal-N$_4$ configurations, namely 
s-MPc and G-MN$_4$, whereas Figures~\ref{f1}(c) and \ref{f1}(d) depict the metal-O$_4$ systems M$_4$(OHTPT)$_2$ and 
M$_3$(HHTP)$_2$, respectively. The stability of the single-layer structures was evaluated by calculating the binding energy of 
the metal center.
Our results reveal that all systems exhibit substantial metal binding energies, ranging from -2.6 to -7.6 eV, largely governed by 
the metal species, indicating high structural stability arising from the interplay between ionic character and covalent orbital overlap. 
Among them, Cu and Zn display the weakest binding, consistent with their fully occupied $3d$ orbitals and indicative of comparatively 
lower thermodynamic stability. Table~\ref{t1} summarizes the calculated structural and electronic properties of the metal-N$_4$ and 
metal-O$_4$ systems. 

We further computed the surface work function ($\Phi$), defined as the minimum energy required to remove an electron from the 
material surface to the vacuum level. This parameter is relevant for assessing the oxidation potential of the metal center, particularly 
in light of the unpaired electrons predominantly localized at the metal site, as discussed below. We find that $\Phi$ for G-MN$_4$ 
is notably lower than in the other structures, suggesting a greater propensity for electron removal in this MN$_4$ coordination
environment.

Figures~S1 and S2 present the calculated band structures of the metal-N$_4$ systems G-MN$_4$ and s-MPc. The G-MN$_4$ 
structures exhibit semiconducting behavior with nearly identical band-gap energies for all metal studied, approximately 0.23~eV. 
The systems G-MnN$_4$, G-FeN$_4$, and G-CoN$4$ display magnetic moments of 3, 2, and 1~$\mu_{\rm B}$, respectively. Their 
magnetization is reflected in a small spin splitting of the bands. In contrast, G-NiN$_4$, G-CuN$_4$, and G-ZnN$_4$ are nonmagnetic. 
For the s-MPc structures, both semiconducting and metallic behaviors are observed depending on the metal atom. The systems 
s-MnPc, s-FePc, and s-CoPc exhibit metallic and magnetic character, with magnetic moments of 3, 2, and 1~$\mu_{\rm B}$, 
respectively. In contrast, s-NiPc, s-CuPc, and s-ZnPc are nonmagnetic semiconductors with band gaps of approximately 0.22~eV. 
The metallic behavior is attributed to significant spin splitting near the Fermi level.

Figures~S3 and S4 present the calculated band structures of the metal-O$_4$ systems M$_4$(OHTPT)$_2$ and M$_3$(HHTP)$_2$. 
Both MOFs exhibit metallic behavior, although their magnetic properties depend on the metal atom. The systems Ni$_4$(OHTPT)$_2$, 
Cu$_4$(OHTPT)$_2$, Zn$_4$(OHTPT)$_2$, and Ni$_3$(HHTP)$_2$ are nonmagnetic, whereas the remaining structures display finite 
magnetic moments. Notably, the magnetic systems exhibit relatively large magnetizations, which can be attributed to metal-oxygen 
hybridization. These systems also show fractional magnetic moments and significant spin polarization on the organic ligands, indicating 
a delocalized magnetic character. This behavior is consistent with the spin-density maps shown in Figures~S5-S8 and with previous 
reports in the literature.\cite{ni2024half} Table~\ref{t1} summarizes the structural, electronic, and magnetic properties of metal-N$_4$  
and metal-O$_4$ systems.
\begin{table}[!htb]
\caption{Structural, electronic and magnetic properties of the metal-N$_4$ and metal-O$_4$ organometallic structures. Lattice constant 
($a$, in \AA), M--X bond distance (X = N, O) ($d_{\rm{M-X}}$, in \AA), work function ($\Phi$, in eV), total magnetic moment ($m$, in 
$\mu_{\rm{B}}$), metal binding energy ($E_{b}$, in eV), and the electronic band gap ($E_{g}$, in eV).}
\begin{ruledtabular}
\begin{tabular}{lccccccc}
System  & $a$   & $d_{\rm M-X}$  & $\Phi$  & $m$ & $E_{b}$  & $E_{g}$ \\ 
\hline
G-MnN$_4$                 & 19.84 & 1.96 & 4.15 & 3.0 & -4.27  & 0.23 \\
G-FeN$_4$                  & 19.84 & 1.95 & 4.16 & 2.0 & -5.28  & 0.22 \\  
G-CoN$_4$                  & 19.84 & 1.92 & 4.17 & 1.0 & -5.35  & 0.23 \\
G-NiN$_4$                   & 19.84 & 1.91 & 4.20 & 0.0 & -5.67  & 0.22 \\ 
G-CuN$_4$                  & 19.84 & 1.96 & 4.28 & 0.0 & -3.60  & 0.24 \\
G-ZnN$_4$                   & 19.84 & 1.97 & 4.20 & 0.0 & -2.63  & 0.24 \\ \\
s-MnPc                         & 10.72 & 1.98  & 4.43 & 3.0 & -6.22 & 0.0  \\
s-FePc                          & 10.70 & 1.97  & 5.20 & 2.0 & -7.58 & 0.0  \\  
s-CoPc                          & 10.68 & 1.95  & 5.04 & 1.0 & -7.34 & 0.0  \\
s-NiPc                           & 10.62 & 1.92  & 5.18 & 0.0 & -7.36 & 0.25 \\ 
s-CuPc                          & 10.68 & 1.97  & 5.24 & 1.0 & -5.64 & 0.22 \\
s-ZnPc                          & 10.73 & 1.99  & 5.13  & 0.0 & -4.29 & 0.21  \\ \\
Mn$_4$(OHTPT)$_2$  & 22.12 &  1.91 & 5.01 & 12.8 & -6.06 & 0.0\\
Fe$_4$(OHTPT)$_2$   & 22.04 &  1.88 & 5.40 & 8.5  & -6.01 & 0.0\\  
Co$_4$(OHTPT)$_2$  & 21.98 &  1.87  & 5.41 & 4.4  & -5.81 & 0.0\\
Ni$_4$(OHTPT)$_2$   & 21.95 &  1.87  & 5.62 & 0.0  & -5.69 & 0.0\\ 
Cu$_4$(OHTPT)$_2$  & 22.03 &  1.94  & 5.43 & 0.0  & -2.70 & 0.0\\
Zn$_4$(OHTPT)$_2$  & 22.07 &  2.00  & 5.11 & 0.0  & -3.75  & 0.0 \\ \\
Mn$_3$(HHTP)$_2$    & 21.54 & 1.89  & 4.98 & 9.0  & -7.16 & 0.0\\
Fe$_3$(HHTP)$_2$     & 21.73 & 1.86  & 5.18 & 6.0  & -7.13 & 0.0\\  
Co$_3$(HHTP)$_2$    & 21.49 & 1.86  & 5.17 & 3.0  & -7.78 & 0.0\\
Ni$_3$(HHTP)$_2$     & 21.44 & 1.85  & 5.02 & 0.0  & -6.57 & 0.0\\ 
Cu$_3$(HHTP)$_2$    & 21.69 & 1.95  & 4.97 & 4.7  & -4.01 & 0.0\\
Zn$_3$(HHTP)$_2$    & 21.72 & 2.00  & 4.83 & 2.0  & -3.18 & 0.0\\ 
\end{tabular}   
\end{ruledtabular}
\label{t1} 
\end{table}

\subsection{Eletrochemical Stability}

The oxygen reduction and evolution reactions are intrinsically sluggish due to their multistep proton-coupled electron transfer mechanisms, 
which impose significant kinetic barriers and lead to substantial overpotentials. Consequently, the design of catalysts that can simultaneously 
enhance activity and maintain long-term stability remains a major challenge. This is particularly critical under harsh operating conditions, 
where acidic and alkaline environments can induce catalyst degradation through dissolution, surface reconstruction, or poisoning effects. 
The electrochemical stability of the layered organometallic structures in aqueous environments is evaluated through the construction 
of surface Pourbaix diagrams, which describe the dependence of the electrode potential on pH. Stability under ORR and OER conditions 
is determined by comparing the relative thermodynamic stability of distinct surface states via the following redox reactions~\cite{jerkiewicz2020}:

\begin{equation}
\text{M}^{z+} + ze^{-} \leftrightharpoons \text{M},
\label{reaction1}
\end{equation}
\begin{equation}
\text{M} + \text{H}^{+} + e^{-} \leftrightharpoons \text{H}^{*},
\label{reaction2}
\end{equation}
\begin{equation}
\text{OH}^{*} + \text{H}^{+} + e^{-} \leftrightharpoons \text{M} + \text{H}_{2}\text{O},
\label{reaction3} 
\end{equation}
\begin{equation}
\text{O}^{*} + 2\text{H}^{+} + 2e^{-} \leftrightharpoons \text{M} + \text{H}_{2}\text{O}.
\label{reaction4} 
\end{equation}

Here, $z$ denotes the number of electrons involved in the electrochemical process. Reaction~(1) represents the oxidation of the metal 
center in the pristine system (M). Reactions~(2)--(4) describe coupled proton-electron transfer processes associated with the 
adsorption of H, OH, and O species on the metal site, respectively. The standard electrode potentials, referenced to the standard hydrogen 
electrode (SHE), were evaluated within the computational hydrogen electrode (CHE) model.~\cite{norskov2004} The explicit consideration 
of H, OH, and O adsorbed species in surface Pourbaix diagrams captures the fundamental proton-electron coupled surface states that govern 
electrochemical stability in aqueous environments. Electrochemical stability can be further assessed by analyzing the energetic relationship 
between the OH$^*$ adsorption potential and the metal oxidation potential. This relative alignment provides a key descriptor for determining 
whether surface oxidation is thermodynamically favored. Under standard conditions (298~K) and applying the Nernst equation, 
the oxidation potential of the metal species can be expressed as:~\cite{dobrota2022}
\begin{equation}
E(\text{M}^{z+}) = E^{\circ}(\text{M}^{z+}) - \frac{0.059}{z} \log a(\text{M}^{z+}),
\label{eq5}
\end{equation}
where $E$ denotes the electrode potential of the electrochemical cell, $E^{\circ}$ represents the standard oxidation potential, and 
$a(\text{M}^{z+})$ the activity of the dissolved metal ions. In the present work, the ionic activity is fixed at 
$a(\text{M}^{z+}) = 1\times10^{-8}$ mol~dm$^{-3}$ and the first oxidation potential ($z = 1$) is assumed.

The first oxidation potential of the coordinated metal centers, $E^{\circ}(\text{M}^{1+})$, was derived from the calculated electronic work 
functions of the organometallic surfaces, under the assumption that electron removal occurs primarily from unpaired electrons localized 
at the coordinated metal sites.\cite{orellana2024,orellana2025,hoyos-sinchi2025} Specifically, $E^{\circ}(\text{M}^{1+})$ was obtained by 
exploiting the linear relationship between metal work functions and their corresponding standard redox potentials, as reported by 
Li et al.\cite{li2022theoretical} Furthermore, within the same theoretical framework, the oxidation potential associated with the key OH$^{*}$
intermediate can be expressed as
\begin{equation}
E(\text{OH}^{*}) = E^{\circ}(\text{OH}^{*}) - 0.059 \times {\rm pH} ,
\label{eq6}
\end{equation}
where $E^{\circ}(\text{OH}^{*})=\Delta G_{\text{OH}^{*}}/eF$, with $\Delta G_{\text{OH}^{*}}$ representing the adsorption free energy of 
the OH$^{*}$ intermediate on the coordinated metal site and $F$ the Faraday constant. 

According to Pourbaix diagrams, the electrochemical stability of organometallic surface catalysts can be assessed by examining the 
dependence of electrode potentials on pH. In this framework, both the electrode potential of the pristine surface, $E(\text{M}^{1+})$, 
and that of the surface with OH adsorbates on the metal sites, $E(\text{OH}^{*})$, are considered. Comparing these potentials as a 
function of pH enables the evaluation of the stability of metal sites against dissolution, which occurs when $E(\text{M}^{1+})$ is lower 
than $E(\text{OH}^{*})$ across the entire pH range.\cite{dobrota2022,orellana2025}

\begin{figure}[t]
\includegraphics[width=7.0cm]{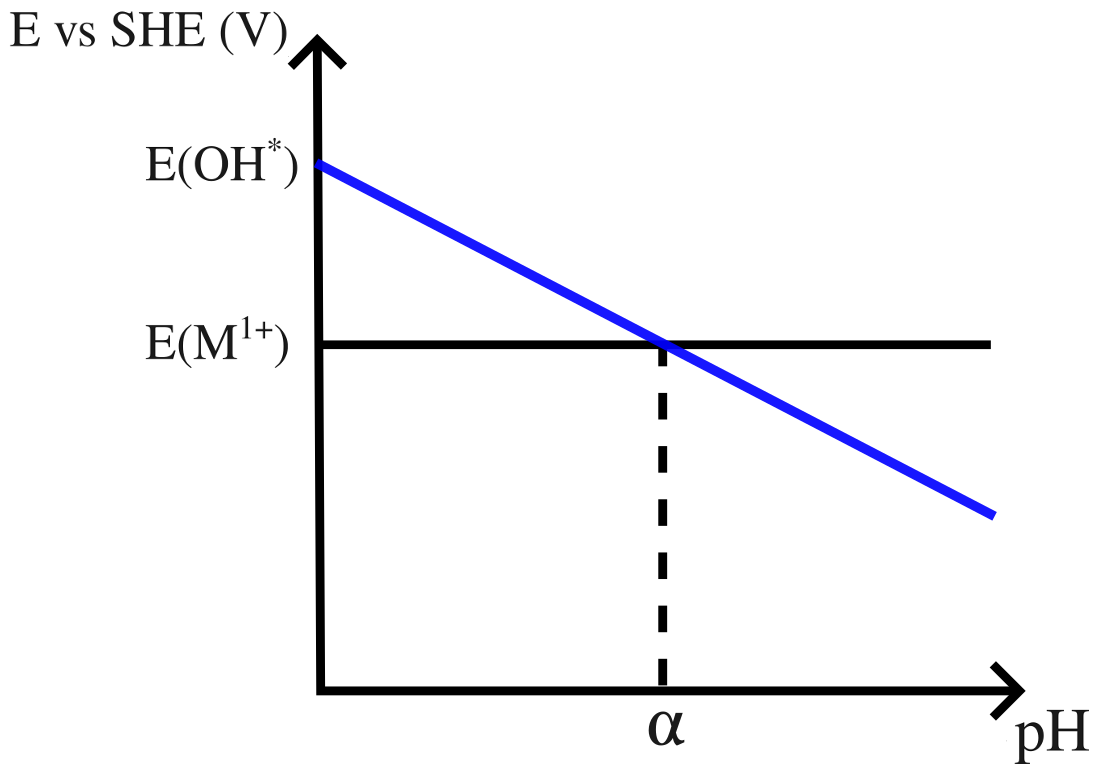}
\caption{Squematic representation of the electrode potential as a function of the pH in a Pourbaix diagram. $E({\rm M}^{1+}$) is the
first oxidation potential for the metal center in the pristine surface. $E({\rm OH}^{*}$) is the  electrode potential of the surface 
with OH adsorbate on the metal sites. $\alpha$ represents the pH at which these two potentials intersect.}
\label{f2}
\end{figure}

Figure~\ref{f2} schematically represent the relative alignment between $E(\text{M}^{1+})$ and $E(\text{OH}^{*})$ in the surface Pourbaix 
diagram. We define $\alpha$ as the pH value at which these two potentials intersect. Therefore, from equations \ref{eq5} and \ref{eq6} 
we obtain the relation
\begin{equation}
\alpha = \frac{D}{\rm 0.059~V} + 8,
\label{eq_alpha}
\end{equation}
where $D=E^{\circ}(\text{OH}^{*})-E^{\circ}(\text{M}^{1+})$. Within this framework, the pH range $0 \le \alpha \le 14$ corresponds to 
$0.5 \le D \le 1.3$~V. Specifically, for $D < 0.5$~V, the systems are electrochemically stable, whereas for $D > 1.3$~V they are unstable. 
For intermediate values, $0.5 \le D \le 1.3$~V, the systems exhibit partial stability depending on the pH. Therefore, we propose $D$ as 
a descriptor of electrochemical stability based on standard oxidation potentials, enabling direct comparison of stability trends across 
different organometallic materials.

\begin{table}[!htb]
\caption{Standard oxidation potential of the coordinated metal center [$E^{0}(\rm{M}^{+})$ vs SHE, in V] and after  the OH adsorption 
[$E^{0}(\rm{OH}^{*})$ vs SHE, in V], together with the electrochemical stability descriptor ($D$ in V).} 
\begin{ruledtabular}
\begin{tabular}{lcccc}
System & $E^{0}(\text{OH}^{*})$ & $E^{0}(\text{M}^{+})$  & $D$ \\ 
\hline
G-MnN$_4$ &-0.55 & -0.85 & 0.30 \\
G-FeN$_4$  &  -0.09 & -0.83 & 0.74 \\ 
G-CoN$_4$  & 1.12 & -0.82 & 1.94 \\
G-NiN$_4$   & 1.37 & -0.78 & 2.15  \\
G-CuN$_4$  & 0.50 & -0.67 & 1.17  \\
G-ZnN$_4$  &  0.14 & -0.78 & 0.92 \\ \\

s-MnPc    & 0.24 & -0.48 & 0.72 \\
s-FePc     & 1.16 &  0.53 & 0.63 \\ 
s-CoPc    & 1.34 &  0.32 & 1.02 \\
s-NiPc     & 1.86 & 0.51 & 1.35 \\
s-CuPc    & 2.25 & 0.58 & 1.67 \\
s-ZnPc    & 1.46 & 0.44 &  1.02 \\ \\
   
Mn$_4$(OHPTP)$_2$  & 0.71 & 0.28 & 0.43 \\
Fe$_4$(OHPTP)$_2$   & 1.72 & 0.79 & 0.93 \\ 
Co$_4$(OHPTP)$_2$  & 0.75 & 0.81 &-0.06 \\
Ni$_4$(OHPTP)$_2$   & 1.51 & 1.08 & 0.43 \\
Cu$_4$(OHPTP)$_2$  & 0.01 & 0.83 & -0.82 \\
Zn$_4$(OHPTP)$_2$  & 0.65 & 0.41 & 0.24 \\ \\

Mn$_3$(HHTP)$_2$  & 0.34 & 0.24 & 0.10 \\
Fe$_3$(HHTP)$_2$   & 1.25 & 0.51 & 0.75 \\  
Co$_3$(HHTP)$_2$   & -0.30 & 0.49 & -0.79 \\
Ni$_3$(HHTP)$_2$    & 1.62 & 0.30 & 1.33 \\ 
Cu$_3$(HHTP)$_2$   & 0.96 & 0.23 & 0.73 \\
Zn$_3$(HHTP)$_2$   & 0.44 & 0.05 &  0.39 \\ 
\end{tabular}   
\end{ruledtabular}
\label{t2} 
\end{table}

Table~\ref{t2} summarizes the calculated $D$ values for the metal-N$_4$ and metal-O$_4$ coordination motif considered in this study. 
For G-MN$_4$, we find that only the system with M = Mn remains stable across the entire pH range. In contrast, for M = Fe, Cu, 
and Zn, the structures exhibit partial stability depending on the pH, whereas for M = Co and Ni they are unstable throughout. These 
findings are consistent with the Pourbaix diagrams presented in Figures~S9. For the s-MPc structures, none of the metal centers 
display stability over the full pH range. Systems with M = Mn, Fe, Co, and Zn are only partially stable depending on the pH, while 
those with M = Ni and Cu are unstable, as illustrated in Figure~S10.

In the case of metal-O$_4$ structures, for M$_4$(OHPTP)$_2$ we observe that for M = Mn, Ni, Cu, Co, and Zn are stable 
across the entire pH range. The only exception is M = Fe, which is stable only for pH $> 8$, as shown in Figures~S11. Finally, 
for the M$_3$(HHTP)$_2$ structures, only the Ni-based system is unstable. The Mn and Cu systems are stable for pH $ > 4$, 
while Co and Zn systems remain stable across the entire pH range, as shown in Figure~S12. Overall, a clear trend emerges: 
metal-N$_4$ structures tend to be less electrochemically stable than their metal-O$_4$ counterparts, suggesting that oxygen 
coordination enhances the electrochemical stability of the metal centers, suppressing metal oxidation over a broad pH range. 
In contrast, nitrogen-coordinated systems tend to exhibit either partial stability only above  a critical pH, or complete instability, 
highlighting the influence of the metal-ligand chemical environment on electrochemical robustness.

We also note that other oxidizing species, such as O$^*$, exhibit higher oxidation potentials than OH$^*$, making the latter the 
primary adsorbate competing with the metal oxidation potential of the pristine surface in contact with the solution. Additionally, 
reduction via H$^*$ adsorption on the metal center is also taken into account in the Pourbaix diagrams shown in Figures~S9-S11.

\subsection{ORR and OER catalytic activity}

The initial step of the ORR process involves the adsorption and activation of O$_2$ at the catalyst active site. Therefore, we 
first analyze the stability and electronic properties associated with O$_2$ adsorption on the metal centers of the metal-N$_4$ 
and metal-O$_4$ systems. The equilibrium geometries, binding energies, and spin-coupling characteristics are summarized in 
Table~S1. Our results show that the binding energies for the metal-O$_4$ centers range from -0.3 to -1.6~eV, indicating a 
moderate interaction between O$_2$ and the metal sites that is strong enough to activate the molecule while avoiding overly 
strong adsorption. In contrast, the metal-N$_4$ systems exhibit binding energies in the narrower range of -0.1 to -0.8~eV, 
shifted toward weaker O$_2$ adsorption relative to their oxygen-coordinated counterparts, suggesting comparatively lower 
catalytic activity.
Furthermore, systematic changes in the magnetic ground state are observed upon O$_2$ adsorption. For M = Ni, Cu, and Zn, 
both metal-N$_4$ and metal-O$_4$ systems adopt high-spin configurations, with the adsorbed O$_2$ stabilized in a triplet state. 
In contrast, for M = Mn, Fe, and Co, the systems favor low-spin configurations, with the adsorbed O$_2$ stabilized in a singlet state.

Following O$_2$ adsorption and activation, the catalytic activity toward the four-electron ORR pathway was evaluated within the 
computational hydrogen electrode (CHE) framework.\cite{norskov2004} In this approach, the reaction leads to the formation of 
hydroxide ions as ${\rm O}_{2} + 2\,{\rm H_{2}O} + 4 e^{-} \longrightarrow 4\,{\rm OH}^{-}$.\cite{shilongli2024} Accordingly, the 
ORR mechanism is described by a sequence of four elementary electrochemical steps involving the successive generation of 
OH$^-$ species.

\begin{equation}
{\rm O}_{2}^{*} + {\rm H_{2}O}(l) + e^{-}  \leftrightharpoons {\rm OOH}^{*} + {\rm OH}^{-},
\end{equation}
\begin{equation}
{\rm OOH}^{*} + e^{-}  \leftrightharpoons {\rm O}^{*} + {\rm OH}^{-},
\end{equation}
\begin{equation}
{\rm O}^{*} + {\rm H_{2}O}(l) + e^{-} \leftrightharpoons {\rm OH}^{*} +  {\rm OH}^{-},
\end{equation}
\begin{equation}
{\rm OH}^{*} + e^{-}  \leftrightharpoons {\rm MOF} + {\rm OH}^{-}.
\end{equation}

For the OER, the reactions proceed in reverse. To assess the energetics of the ORR/OER reactions, we plotted adsorption 
free energy profiles for the reaction intermediates OOH$^{*}$, O$^{*}$, and OH$^{*}$, each bound to the metal cation of systems 
studies.  

At zero applied potential referenced to the reversible hydrogen electrode (RHE), i.e., $U = 0~\rm{V}$, all elementary reaction steps 
are thermodynamically downhill. When the potential is increased to the standard equilibrium value of $U = 1.23~\rm{V}$, the reaction 
free energies are shifted downward in proportion to the number of electrons involved in each elementary step. The minimum applied 
potential at which every step becomes energetically favorable defines the onset potential, denoted as $U_{\rm{onset}}$.
The theoretical overpotential, $\eta$, which serves as a key metric for ORR/OER electrocatalytic performance, is determined from 
the deviation of the onset potential from the equilibrium value. Accordingly, the overpotentials for ORR and OER are given by 
$\eta_{\rm{ORR}} = 1.23~\rm{V} - U_{\rm{onset}}$ and $\eta_{\rm{OER}} = U_{\rm{onset}} - 1.23~\rm{V}$, respectively. A detailed 
description of the computational procedure employed to evaluate the adsorption free energies of reaction intermediates and the 
associated ORR/OER overpotentials is provided in the Supplemental Material.

\begin{table}[!htb]
\caption{Catalytic activity descriptors for the single-layer  organometallic structures, including the $d$-band center ($\varepsilon_d$, 
in eV) relative to the Fermi energy and  ORR/OER overpotentials ($\eta$, in V). Previously reported theoretical overpotential values 
for ORR on Pt(111)\cite{hansen2008} and OER on IrO$_2$(110)\cite{xu2018}  are shown for reference.}
\begin{ruledtabular}
\begin{tabular}{lccccc}
System & $\varepsilon_d$ & $\eta_{\rm{ORR}}$  & $\eta_{\rm{OER}}$   \\ 
\hline
G-MnN$_4$ &-0.80& 1.78 &  2.51 \\
G-FeN$_4$ &-1.01&  1.32 &  2.57 \\  
G-CoN$_4$ &-2.01&   0.41 &1.55 \\
G-NiN$_4$ &-2.71&   0.76 & 0.62 \\ 
G-CuN$_4$ &-4.55&  0.73 & 1.50 \\
G-ZnN$_4$ &-7.68&  1.09 & 1.84 \\ \\

s-MnPc &-1.43&  0.98 & 1.80 \\
s-FePc &-0.88&  0.27 & 1.05 \\  
s-CoPc &-1.82& 0.38 &  0.85 \\
s-NiPc  &-2.79&  0.92 & 1.00 \\ 
s-CuPc &-4.50& 0.91 &  0.75 \\
s-ZnPc &-7.53&  0.82 & 0.88 \\ \\
 
Mn$_4$(OHPTP)$_2$ &-1.23&   1.14 &  1.51 \\
Fe$_4$(OHPTP)$_2$ &-1.72&  1.65 &  0.85 \\  
Co$_4$(OHPTP)$_2$ &-2.02&   0.48 &  0.61 \\
Ni$_4$(OHPTP)$_2$ &-1.98&   1.11 &  0.84 \\ 
Cu$_4$(OHPTP)$_2$ &-3.15&   1.22 &  0.91 \\
Zn$_4$(OHPTP)$_2$ &-6.86&   0.58 &  0.55 \\ \\

Mn$_3$(HHTP)$_2$ &-0.93&   0.89 &  0.91\\
Fe$_3$(HHTP)$_2$ &-1.17&  0.51 &  0.25  \\  
Co$_3$(HHTP)$_2$ &-1.28&  1.53 &  1.04  \\
Ni$_3$(HHTP)$_2$ &-1.56&   0.48 &  0.39  \\ 
Cu$_3$(HHTP)$_2$ &-2.64&   1.14 &  0.82  \\
Zn$_3$(HHTP)$_2$ &-5.74&  0.87 &  1.37   \\ \\

Pt(111)           &-&  0.48  & - \\
IrO$_2$(110) &-&    -      &  0.65 \\
\end{tabular}   
\end{ruledtabular}
\label{t3} 
\end{table}

Table~\ref{t3} summarizes the calculated ORR and OER overpotentials for the metal-N$_4$ and metal-O$_4$ structures, 
together with the $d$-band center, which is used as a descriptor of catalytic activity. Previously reported ORR and OER 
overpotentials for Pt(111) and IrO$_2$(110) surfaces, obtained using a similar theoretical approach, are included as 
benchmarks.\cite{hansen2008,xu2018} Figures~S13--S20 present the corresponding ORR and OER  free energy diagrams 
for all the structures considered. To further corroborate our results, we calculated the $d$-band center for all systems, 
confirming that the highest ORR/OER activity is generally associated with $d$-band centers closer to the Fermi energy, 
as shown in Figures~S21--S24.

Among the metal-N$_4$ systems, G-CoN$_4$ and G-NiN$_4$ exhibit the best catalytic performance for ORR and OER, 
with overpotentials of 0.41 and 0.62~V, respectively, comparable to the benchmark values. For ORR, our results are in 
good agreement with available experimental data.\cite{nali2023} In the case of s-MPc structures, the most active catalysts 
for ORR are s-FePc and s-CoPc, with overpotentials of 0.27 and 0.38~V, respectively, whereas none of these structures 
exhibit OER overpotentials lower than the benchmark. Experimental studies on Co-phthalocyanine-based MOFs report 
ORR activities approaching those of Pt/C benchmark catalysts,\cite{shantharaja2023} and also superior activity for 
OER,\cite{kim2021} which agree well with our predictions.

For the metal-O$_4$ systems, Co$_4$(OHPTP)$_2$ exhibits good catalytic activity for both ORR and OER, with 
overpotentials of 0.48 and 0.61~V, respectively, while Zn$_4$(OHPTP)$_2$ also shows a relatively low OER overpotential 
of 0.55~V. Notably, both structures display bifunctional behavior. In addition, Fe$_3$(HHTP)$_2$ and Ni$_3$(HHTP)$_2$ 
exhibit ORR activity comparable to benchmark values, while delivering significantly improved OER performance. 
Experimental data are available for M$_3$(HHTP)$_2$ systems; in particular, Co$_3$(HHTP)$_2$ and Ni$_3$(HHTP)$_2$ 
have been demonstrated to be efficient OER catalysts with enhanced stability,\cite{zhang2018,dong2024} in agreement 
with our predictions.

Concerning electrochemical stability, a growing body of experimental evidence indicates that G-FeN$_4$ centers, 
although active for the oxygen reduction reaction, are intrinsically prone to degradation under operating 
conditions.\cite{shao2019,xia2021,gorgen2022} Recent studies have shown progressive degradation due to reactive
oxygen species attack on the active site, inducing oxidative Fe demetallation and the concomitant oxidation of the 
surrounding carbon matrix.\cite{menga2024,xie2025,chen2025,kobayashi2025} Similar degradation pathways have 
been reported for other metal centers, including G-CoN$_4$ and G-MnN$_4$,\cite{li2018,nali2023,yuanli2024} showing 
metal-dependent stability trends. Collectively, these findings identify demetalation and support oxidation as the primary 
mechanisms limiting the durability of G-MN$_4$ catalysts, underscoring an inherent trade-off between catalytic activity 
and electrochemical stability.\cite{gorgen2022} 

\begin{figure}[!h]
\includegraphics[width=8.5cm]{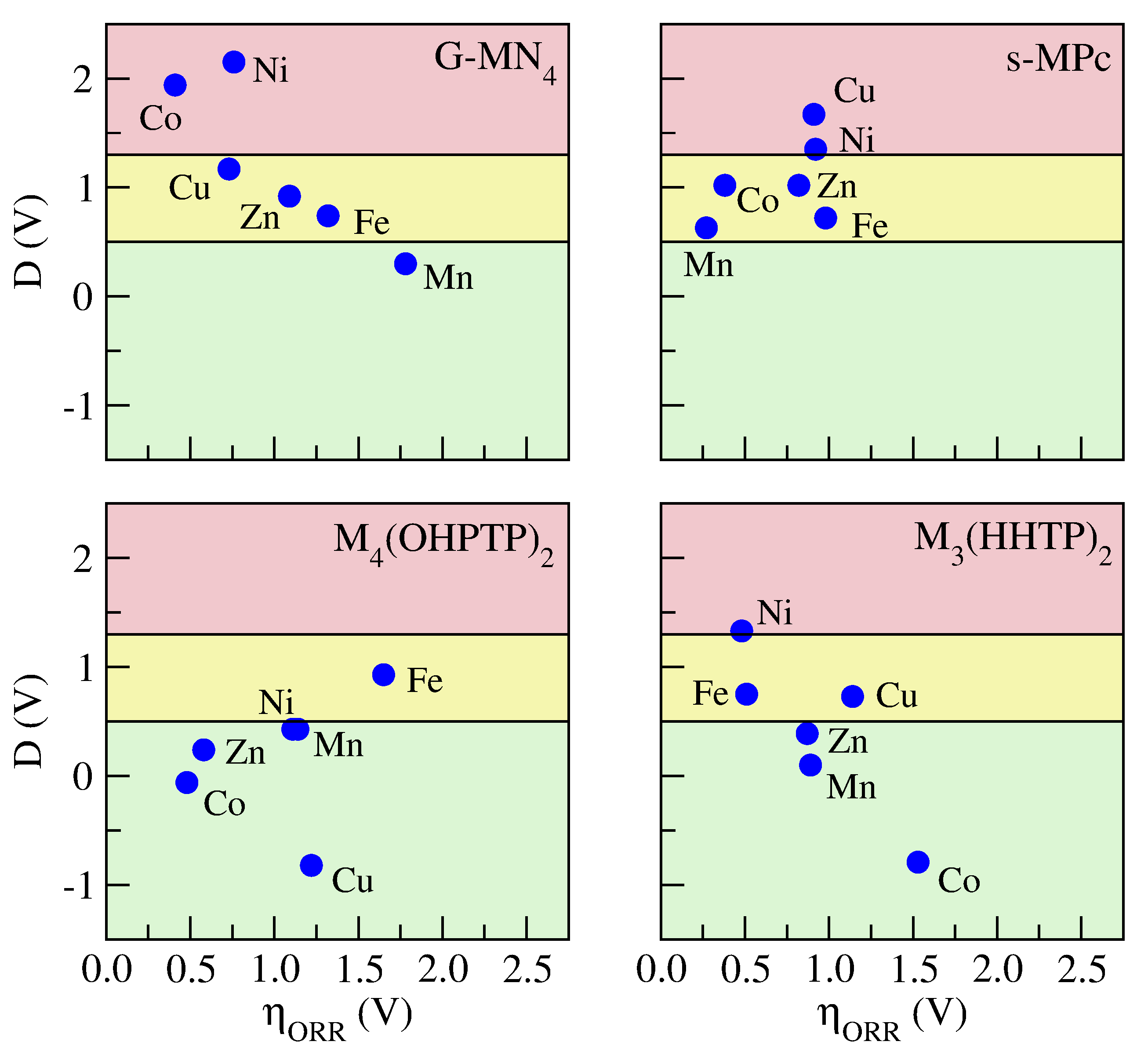}
\caption{Electrochemical stability descriptor ($D$) as a function of the ORR overpotential ($\eta_{\rm{ORR}}$) for the 
organometallic structures. The red, yellow, and green regions denote electrochemically unstable, partially stable, and 
stable regimes, respectively.}
\label{f3}
\end{figure}

\begin{figure}[!h]
\includegraphics[width=8.5cm]{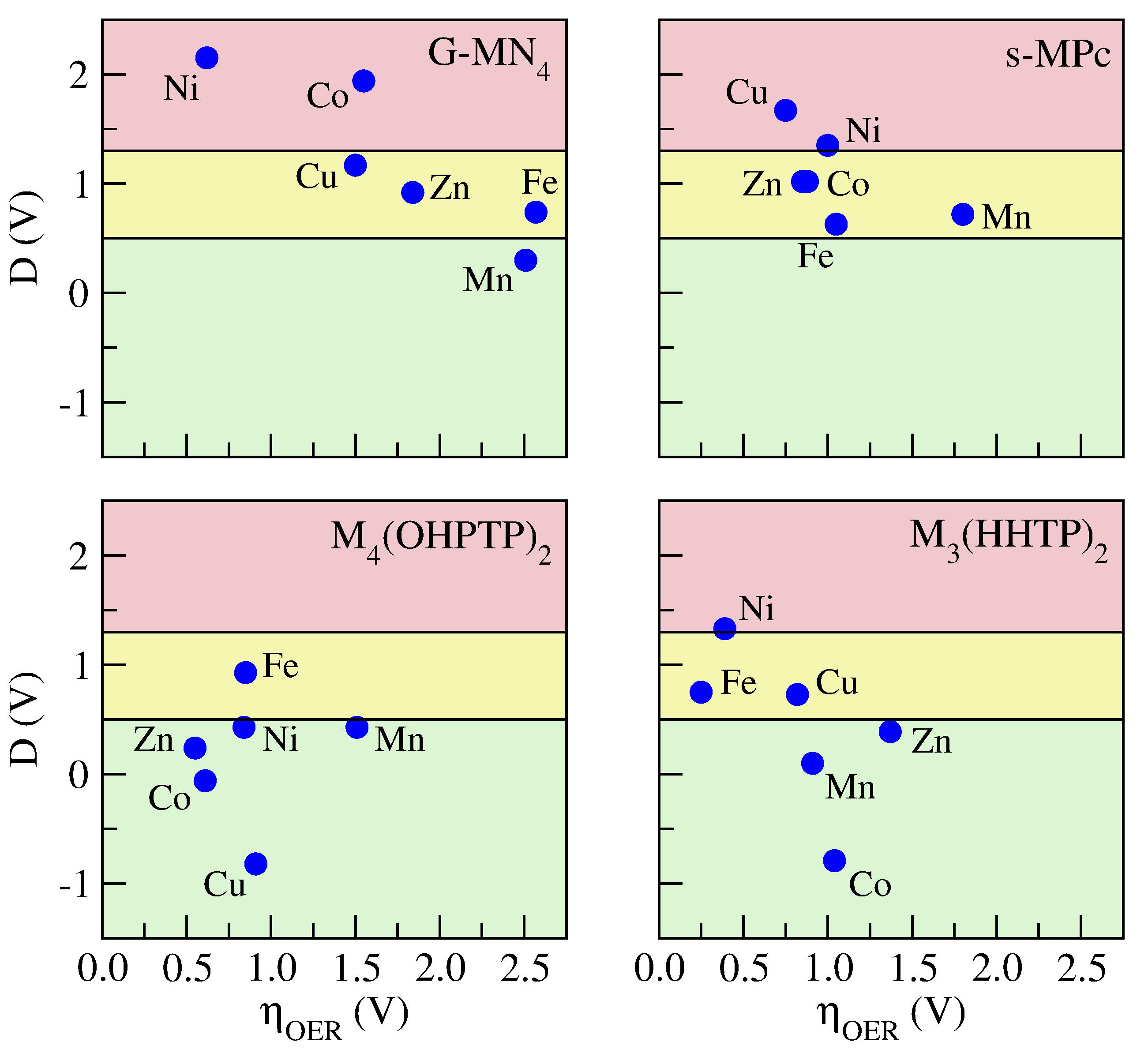}
\caption{Electrochemical stability descriptor ($D$) as a function of the OER overpotential ($\eta_{\rm{OER}}$) for the 
organometallic structures. The red, yellow, and green regions denote electrochemically unstable, partially stable, and 
stable regimes, respectively.}
\label{f4}
\end{figure}

To identify metal-N$_4$ and metal-O$_4$ structures that simultaneously satisfy catalytic activity and electrochemical 
stability, and to enable a direct comparison between them, we plot the stability descriptor $D$ against the ORR and OER 
overpotentials. Figure~\ref{f3} and \ref{f4} present the $D$ descriptor as a function of ORR and OER overpotentials, respectively, 
for the systems under study. The three colored regions of the figures, red, yellow, and green, represent electrochemically 
unstable, partially stable, and stable regimes, respectively. As observed, metal-N$_4$ structures are considerably less 
stable than their metal-O$_4$ counterparts for most metal centers, suggesting a higher susceptibility to degradation. This 
behavior is consistent with experimental observations reported for G-MN$_4$ catalysts. In contrast, none of the 
metal-O$_4$ structures fall within the instability regime. 

Therefore, based on the combined analysis of catalytic activity and electrochemical stability across the four investigated 
systems, oxygen-coordinated metal frameworks emerge as the most promising candidates for practical electrocatalytic 
applications. These systems consistently exhibit a greater tendency toward electrochemical stability over a broad pH 
range, while maintaining competitive overpotential values for both the ORR and OER. In contrast, although nitrogen-coordinated 
metal systems often display favorable overpotentials, indicative of strong intrinsic catalytic activity, our results suggest a 
higher susceptibility to degradation, in good agreement with experimental reports on G-MN$_4$ structures. Such limitations 
include restricted stability depending on pH and, in some cases, complete electrochemical instability, which may ultimately 
hinder their practical applicability under operational conditions.

\section{Summary and Conclusions}

In summary, DFT calculations were employed to investigate the electrochemical stability and  ORR/OER catalytic activity 
of layered organometallic materials. The studied systems comprise metal-N$_4$ structures: G-MN$_4$ and s-MPc, and 
metal-O$_4$ structures: M$_4$(OHPTP)$_2$ and M$_3$(HHTP)$_2$, with M = Mn, Fe, Co, Ni, Cu, and Zn. ORR and 
OER activities were evaluated using the CHE model, while electrochemical stability was assessed through surface 
Pourbaix analysis. To quantify electrochemical stability, we introduce the descriptor $D$, defined from the standard 
potentials, which is obtained from the electronic work function of the pristine surface and the corresponding OH-covered 
surface at the metal active site. This descriptor enables a direct comparison of stability across different metal coordination 
environments.

Our results reveal that metal-N$_4$ systems often exhibit favorable overpotentials but suffer from pronounced pH-dependent 
stability limitations. In contrast, metal-O$_4$ systems display enhanced electrochemical robustness while maintaining competitive 
ORR and OER performance, thereby offering a more balanced combination of activity and stability. Particularly noteworthy are 
the Co$_4$(OHPTP)$_2$ and Zn$_4$(OHPTP)$_2$ structures, which exhibit both high stability and bifunctional ORR/OER 
activity. Similarly, Fe$_3$(HHTP)$_2$ shows excellent ORR/OER activity, although it becomes unstable under highly acidic 
conditions. Finally, owing to its clear physical meaning and quantitative nature, the proposed $D$ descriptor serves as a 
transferable metric for screening the electrochemical stability of layered organometallic catalysts.

\section{Acknowledgments}
We acknowledge financial support from the National Agency for Research and Development of Chile (ANID), 
through projects FONDECYT 1230138 and FONDECYT POSTDOC 3240440 (P.H.S.). The computational 
resources were provided by Fenix HPC (UNAB) and the supercomputing infrastructure of the NLHPC (CCSS210001).

\bibliography{mybibfile}

@article{yang2021,
  title={Regulating {Fe}-spin state by atomically dispersed {Mn-N} in {Fe-NC} catalysts with high oxygen reduction activity},
  author={ G. Yang and  J. Zhu and P. Yuan and Y. Hu and G. Qu and B. Lu and others},
  journal={Nat. Commun.},
  volume={12},
  pages={1734},
  year={2021},
  doi={10.1038/s41467-021-21919-5}
}

@article{gong2023,
author = {M. Gong and A. Mehmood and B. Ali and K. Nam and A. Kucernak},
title = {Oxygen Reduction Reaction Activity in Non-Precious Single-Atom ({M-N/C}) Catalysts-Contribution of Metal and Carbon/Nitrogen Framework-Based Sites},
journal = {ACS Catal.},
volume = {13},
number = {10},
pages = {6661-6674},
year = {2023},
doi = {10.1021/acscatal.3c00356},
}

@article{lv2023,
  title={Electrocatalytic porphyrin/phthalocyanine-based organic frameworks: building blocks, coordination microenvironments, structure-performance relationships},
  author={N. Lv and Q. Li and H. Zhu and S. Mu and X. Luo and X. Ren and X. Liu and S. Li and C. Cheng and T. Ma},
  journal={Adv. Sci.},
  volume={10},
  number={7},
  pages={2206239},
  year={2023},
  publisher={Wiley Online Library},
  doi={10.1002/advs.202206239}
}

@article{zhu2019,
title = {{N,P Co}-Coordinated Manganese Atoms in Mesoporous Carbon for Electrochemical Oxygen Reduction},
author = {X. Zhu and R. Amal and X. Lu},
journal = {Small},
volume = {15},
number = {29},
pages = {1804524},
doi = {10.1002/smll.201804524},
year = {2019}
}

@article{kumar2026,
title = {Advances in coordination engineering of {M-N-C} single atom catalysts for superior oxygen reduction performance},
author = {A. Kumar and N. Goyal and S. Mathur and I. A. Bakhtiyarovich and Y. Zhao and M. Khalid and M. Ubaidullah and A. M. Al-Enizi},
journal = {Coord. Chem. Rev.},
volume = {549},
pages = {217244},
year = {2026},
doi = {10.1016/j.ccr.2025.217244}
}

@article{jin2021,
title   = "Understanding the Inter-Site Distance Effect in Single-Atom Catalysts for Oxygen Electroreduction", 
author  = "Z. Jin and P. Li and Y. Meng and Z. Fang and D. Xiao and G. Yu", 
journal = "Nat. Catal.", 
volume  = "4", 
pages   = "615--622", 
year    = "2021",
doi     = "10.1038/s41929-021-00650-w", 
}

@article{zhou2025,
title = {Structural-property relationships of phenyl-based multifunctional hybrid ligand {2D} {MOFs} ({Ni-X$_n$Y$_{4-n}$, where {X, Y = NH, O, S}}): A theoretical study},
author = {J. Zhou and W. Zhao and M. Zhang and A. Zhang and H. Ren and H. Zhu and Y. Chi and W. Guo},
journal = {Appl. Surf. Sci.},
volume = {689},
pages = {162541},
year = {2025},
doi = {10.1016/j.apsusc.2025.162541},
}

@article{meng2009,
title = {{pH}-effect on oxygen reduction activity of {Fe}-based electro-catalysts},
author = {H. Meng and F. Jaouen and E. Proietti and M. Lef\'evre and J. Dodelet},
journal = {Electrochem. Commun.},
volume = {11},
pages = {1986--1989},
year = {2009},
doi = {10.1016/j.elecom.2009.08.035},
}

@article{cwan2020,
title   = "Molecular Design of Single-Atom Catalysts for Oxygen Reduction Reaction",
author  = "C. Wan and X. Duan and Y. Huang",
journal = "Adv. Energy Mater.",
volume  = "10",
pages   = "1903815",
year    = "2020",
doi     = "10.1002/aenm.201903815"
}

@article{jli2021,
title   = "A General Strategy for Preparing Pyrrolic-{N$_4$} Type Single-Atom Catalysts via Pre-Located Isolated Atoms",
author  = "J. Li and Y.-F Jiang and Q. Wang and C.-Q. Xu and D. Wu and M. N. Banis and others",
journal = "Nat. Commun.",
volume  = "12",
pages   = "6806",
year    = "2021",
doi     = "10.1038/s41467-021-27143-5"
}

@article{wei2022no, 
title = {Novel Two-dimensional Metal Organic Frameworks: High-performance Bifunctional Electrocatalysts for {OER/ORR}},
author = {X. Wei and S. Cao and H. Xu and C. Jiang and Z. Wang and Y. Ouyang and X. Lu and F. Dai and D. Sun}, 
journal = {ACS Mater. Lett.}, 
volume = {4},
pages = {1991--1998}, 
year = {2022},
doi = {10.1021/acsmaterialslett.2c00694}
}

@article{gorgen2022,
author = {E. Kolle-G\"orgen and G. Fortunato and M. Ledendecker},
title = {Catalyst Stability in Aqueous Electrochemistry},
journal = {Chem. Mater.},
volume = {34},
pages = {10223--10236},
year = {2022},
doi = {10.1021/acs.chemmater.2c02443}
}

@article{DF2,
  title = "Higher-accuracy van der {Waals} density functional",
  author = " K. Lee and  \'E. D. Murray and  L. Kong and B. I. Lundqvist and  D. C. Langreth",
  journal = "Phys. Rev. B",
  volume = "82",
  issue = "8",
  pages = "081101",
  numpages = "4",
  year = "2010",
  month = "Aug",
  publisher = "American Physical Society",
  doi = "10.1103/PhysRevB.82.081101",

}

@article{hamada2018,
 author = {T. Hamada and T. Ohno},
title = {A new constraint {DFT} technique for self-consistent determination of {U} values},
journal = {J. Phys.: Condens. Matter.},
volume = {31},
pages = {065501},
year = {2018},
doi = {10.1088/1361-648X/aaf6f4}
}

@article{vaspkit,
  title   = "{VASPKIT:} A {User-Friendly} Interface Facilitating {High-Throughput} Computing and Analysis Using {VASP} Code" ,
  author  = "V. Wang and N. Xu and J. Liu and G. Tang and W. Geng" ,
  journal = "Comput. Phys. Commun.",
  volume  = "267" ,
  pages   = "108033" ,
  year    = "2021" ,
  doi     = "10.1016/j.cpc.2021.108033" ,
}

@article{norskov2004,
author = "J. K. N{\o}rskov and J. Rossmeisl and A. Logadottir and L. Lindqvist and J. R. Kitchin and T. Bligaard and H. J\'onsson",
title = "Origin of the Overpotential for Oxygen Reduction at a Fuel-Cell Cathode",
journal = "J. Phys. Chem. B.",
volume = "108",
pages = "17886-17892",
year = "2004",
doi = "10.1021/jp047349j",
}

@article{ni2024half,
  title={Half-Metallic Ferromagnetism in Radical-Bridged Metal-Organic Frameworks},
  author={X. Ni and H. Li and J. L. Bredas},
  journal={Chem. Mater.},
volume = {36},
pages = {2380-2389},
year = {2024},
doi={10.1021/acs.chemmater.3c03039}
}

@article{jerkiewicz2020,
author = {Jerkiewicz, G.},
title = {Standard and Reversible Hydrogen Electrodes: Theory, Design, Operation, and Applications},
journal = {ACS Catalysis},
volume = {10},
pages = {8409--8417},
year = {2020},
doi = {10.1021/acscatal.0c02046}
}

@article{dobrota2022,
author = {A. S. Dobrota and N. V. Skorodumova and Slavko V. Mentus and I. A. Pasti},
title = {Surface Pourbaix Plots of {M@N$_4$}-Graphene Single-Atom Electrocatalysts from Density Functional Theory Thermodynamic Modeling},
journal = {Electrochim. Acta},
volume = {412},
pages = {140155},
year = {2022},
doi = {10.1016/j.electacta.2022.140155},
}

@article{li2022theoretical,
  title={Theoretical relations between electronic and ionic work functions, standard reduction potentials for metal dissolution and the corrosion potential},
  author={Li, S. and Frankel, G. S. and Taylor, C. D.},
  journal={J. Electrochem. Soc.},
  volume={169},
  pages={081506},
  year={2022},
  doi = {10.1149/1945-7111/ac86f8}
}

@article{vesta,
  title   = "{VESTA}: A Three-Dimensional Visualization System for Electronic and Structural Analysis" ,
  author  = "K. Momma and F. Izumi",
  journal = "J. Appl. Crystallogr." ,
  volume  = "41" ,
  pages   = "653--658" ,
  year    = "2008" ,
  doi     = "10.1107/S0021889808012016" ,
}

@article{vasp,
  title="Efficient Iterative Schemes for Ab Initio Total-Energy Calculations Using a Plane-Wave Basis Set",
  journal="Phys. Rev. B",
  author="G. Kresse and J. {Furthm\"uller}",
  volume="54",
  pages="11169",
  year="1996",
  doi="10.1103/PhysRevB.54.11169"
}

@article{pathak2024,
  title={Structural and Phase Engineering of a Hierarchical {2D-2D} Nickel {MOF}/Hydroxide-Derived {Ni$_{0.85}$Se/NiTe$_2$} Heterointerface for Robust {HER}, {OER}, and Overall Water Splitting},
  author={Pathak, I. and Prabhakaran, S. and Acharya, D. and Chhetri, K. and Muthurasu, A. and Rosyara, Y. R. and others},
  journal={Small},
  volume={20},
  pages={2406732},
  year={2024},
  doi={10.1002/smll.202406732} 
}

@article{xiao2022,
  title={{2D} {MOFs} and Their Derivatives for Electrocatalytic Applications: Recent Advances and New Challenges},
  author={Xiao, L. and Wang, Z. and Guan, J.},
  journal={Coord. Chem. Rev.},
  volume={472},
  pages={214777},
  year={2022},
  publisher={Elsevier}, 
  doi={10.1016/j.ccr.2022.214777}
}

@article{xing2025,
  title={Activation of Efficient Hydrogen Evolution on Two-Dimensional Kagome Structured Perthiolated Coronene {(PTC)} Metal-Organic Frameworks {(MOFs)}: A First-Principles Study},
  author={Xing, D. and Jia, M. and Ding, P. and Tao, J. and Zhang, S. and Guan, L.},
  journal={Int. J. Hydrogen Energy.},
  volume={100},
  pages={191--200},
  year={2025},
  doi={10.1016/j.ijhydene.2024.12.315}
}

@article{wang2024,
  title={Two-dimensional Conductive Metal-Organic Frameworks Electrocatalyst: Design Principle and Energy Conversion Applications},
  author={Wang, X. and Borse, R. A. and Wang, G. and Xiao, Z. and Zhu, H. and Sun, Y. and Qian, Z. and Zhong, S. and Wang, R.},
  journal={Mater. Today Energy},
  pages={101652},
  year={2024},
  publisher={Elsevier}, 
  doi={10.1016/j.mtener.2024.101652}
}

@article{li2024,
  title={{2D} Conductive Metal--Organic Frameworks for Electrochemical Energy Application},
  author={Li, R. and Yan, X. and Chen, L.},
  journal={Org. Mater.},
  volume={6},
  pages={45--65},
  year={2024},
  publisher={Georg Thieme Verlag KG},
  doi={10.1055/s-0044-1786500}
}

@article{yang2022,
  title={{2D} {$\pi$-Conjugated} Metal-organic Frameworks for {CO$_2$} Electroreduction},
  author={Yang, D. and Wang, X.},
  journal={SmartMat},
  volume={3},
  pages={54--67},
  year={2022},
  doi={10.1002/smm2.1102}
}

@article{yi2021,
  title={Conductive Two-Dimensional Phthalocyanine-Based Metal-Organic Framework Nanosheets for Efficient Electroreduction of {CO$_2$}},
  author={Yi, J.-D. and Si, D.-H. and Xie, R. and Yin, Q. and Zhang, M.-D. and Wu, Q. and Chai, G.-L. and Huang, Y.-B. and Cao, R.},
  journal={Angew. Chem.},
  volume={133},
  pages={17245--17251},
  year={2021},
  doi={10.1002/ange.202104564}
}

@article{cui2019,
  title={{Mo-based} {2D MOF} as a Highly Efficient Electrocatalyst for Reduction of {N$_2$} to {NH$_3$}: a Density Functional Theory Study},
 author={Cui, Q. and Qin, G. and Wang, W. and Geethalakshmi, K. R. and Du, A. and Sun, Q.},
  journal={J. Mater. Chem. A.},
  volume={7},
  number={24},
  pages={14510--14518},
  year={2019},
  publisher={Royal Society of Chemistry}, 
  doi={10.1039/C9TA02926E}
}

@article{li2020,
  title={Bimetal-{MOF} Nanosheets as Efficient Bifunctional Electrocatalysts for Oxygen Evolution and Nitrogen Reduction Reaction},
  author={Li, W. and Fang, W. and Wu, C. and Dinh, K. N. and Ren, H. and Zhao, L. and Liu, C. and Yan, Q.},
  journal={J. Mater. Chem. A.},
  volume={8},
  number={7},
  pages={3658--3666},
  year={2020},
  publisher={Royal Society of Chemistry}, 
  doi={10.1039/C9TA13473E}
}

@article{li2018,
  title   = "Atomically Dispersed Manganese Catalysts for Oxygen Reduction in Proton-Exchange Membrane Fuel Cells", 
  author  = "J. Li and M. Chen and D.  A. Cullen and S. Hwang and M. Wang and B. Li and others" ,
  journal = "Nat. Catal.",
  volume  = "1",
  pages   = "935--945",
  year    = "2018",
  doi     = "10.1038/s41929-018-0164-8",
}

@article{xu2018,
  title={A Universal Principle for a Rational Design of Single-Atom Electrocatalysts},
  journal={Nat. Catal.},
  author={H. Xu and D. Cheng and D. Cao and X. C. Zeng},
  volume={1},
  pages={339--348},
  year={2018},
  doi={10.1038/s41929-018-0063-z},
}

@article{orellana2024,
  title   = "{Fe} and {Co} Adatoms on Bilayer Borophene as Single-Atom Catalysts for the Oxygen Reduction Reaction: A Theoretical Study", 
  author  = "W. Orellana and R. H. Miwa" ,
  journal = "Phys. Rev. Appl." ,
  volume  = "21" ,
  pages   = "034008" ,
  year    = "2024" ,
  doi     = "10.1103/PhysRevApplied.21.034008" ,
}

@article{mathew2014,
  title={Implicit Solvation Model for Density-Functional Study of Nanocrystal Surfaces and Reaction Pathways},
  journal={J. Chem. Phys.},
  author={K. Mathew and R. Sundararaman and K. Letchworthweaver and T. A. Arias and R. G. Hennig},
  volume={140},
  pages={084106},
  year={2014},
  doi={10.1063/1.4865107},
  publisher={AIP} 
}

@article{orellana2025,
  title={Stability and Activity of Organometallic Phthalocyanine Sheets for Oxygen Reduction and Oxygen Evolution Reactions: A {DFT} Study},
  journal={Electrochim. Acta},
  author={W. Orellana},
  volume={514},
  pages={145602},
  year={2025},
  doi={10.1016/j.electacta.2024.145602},
}

@article{hansen2008,
  title="Surface Pourbaix Diagrams and Oxygen Reduction Activity of {Pt}, {Ag} and {Ni(111)} Surfaces Studied by {DFT}",
  journal="Phys. Chem. Chem. Phys",
  author="H. A. Hansen and J. Rossmeisl and J. K. N{\o}rskov",
  volume="10",
  pages="3722--3730",
  year="2008",
  doi="10.1039/B803956A"
}

@article{souza2025,
  title="Electronic Properties and Stability of Single-Layer and Multilayer {Cu$_3$(HHTP)$_2$} Metal-Organic Frameworks",
  journal="J. Phys. Chem. C",
  author="P. H. Souza and W. Orellana",
  volume="129",
  pages="3285--3291",
  year="2025",
  doi="10.1021/acs.jpcc.4c07540"
}

@article{shilongli2024,
  title="Selective Oxygen Reduction Reaction: Mechanism Understanding, Catalyst Design and Practical Application",
  journal="Chem. Sci.",
  author="S. Li and L. Shi and Y. Guo and J. Wang and D. Liu and S. Zhao",
  volume="15",
  pages="11188--11228",
  year="2024",
  doi="10.1039/D4SC02853H"
}

@article{hongxing2018,
  title="A Novel Two-Dimensional Nickel Phthalocyanine-Based Metal-Organic Framework for Highly Efficient Water Oxidation Catalysis",
  journal="J. Mater. Chem. A",
  author="H. Jia and Y. Yao and J. Zhao and Y. Gao and Z. Luo and P. Du",
  volume="6",
  pages="1188",
  year="2018",
  doi="10.1039/c7ta07978h"
}

@article{mingdao2018,
  title="Fewer-Layer Conductive Metal-Organic Nanosheets Enable Ultrahigh Mass Activity for the Oxygen Evolution Reaction",
  journal="Chem. Commun.",
  author="M. Zhang and B.-H. Zheng and J. Xu and N. Pan and J. Yu and M. Chen and H. Cao",
  volume="54",
  pages="13579",
  year="2018",
  doi="10.1039/c8cc08156e"
}

@article{miner2016,
  title="Electrochemical Oxygen Reduction Catalysed by {Ni$_3$(hexaiminotriphenylene)$_2$}",
  journal="Nat. Commun.",
  author="E. M. Miner and T. Fukushima and D. Sheberla and L. Sun and Y. Surendranath and M. Dinc\u{a}",
  volume="7",
  pages="10942",
  year="2016",
  doi="10.1038/ncomms10942"
}

@article{park2020,
  title="Two-Dimensional Conductive {Ni-HAB} as a Catalyst for the Electrochemical Oxygen Reduction Reaction",
  journal="ACS Appl. Mater. Interfaces",
  author="J. Park and Z. Chen and R. A. Flores and G. {Wallnerstr\"om} and A. Kulkarni and J. K. N{\o}rskov and T. F. Jaramillo and Z. Bao",
  volume="12",
  pages="39074",
  year="2020",
  doi="10.1021/acsami.0c09323"
}

@article{jiawenli2021,
  title="Structural and Electronic Modulation of Conductive {MOFs} for Efficient Oxygen Evolution Reaction Electrocatalysis",
  journal="J. Mater. Chem. A",
  author="J. Li and P. Liu and J. Mao and J. Yan and W. Song",
  volume="9",
  pages="11248",
  year="2021",
  doi="10.1039/d1ta01970h"
}

@article{qianzhao2021,
  title="Truxone-Based Conductive Metal-Organic Frameworks for the Oxygen Reductive Reaction",
  journal="J. Phys. Chem. C",
  author="Q. Zhao and J. Jiang and W. Zhao and S.-H. Li and W. Mi and C. Zhang",
  volume="125",
  pages="12690",
  year="2021",
  doi="10.1021/acs.jpcc.1c03418"
}

@article{wang2022,
  title="Two-Dimensional Metal-Organic Frameworks as Efficient Electrocatalysts for Bifunctional Oxygen Evolution/Reduction Reactions",
  journal="J. Mater. Chem. A",
  author="A. Wang and H. Niu and X. Wang and X. Wan and L. Xie and Z. Zhang and J. Wang and Y. Guo",
  volume="10",
  pages="13005",
  year="2022",
  doi="10.1039/d2ta01319c"
}

@article{xuejingyang2018,
  title="Oxygen Reduction Reaction on {M$_3$}(hexaiminobenzene)$_2$: A Density Function Theory Study",
  journal="Catal. Commun.",
  author="X. Yang and Q. Hu and X. Hou and J. Mi and P. Zhang",
  volume="115",
  pages="17--20",
  year="2018",
  doi="10.1016/j.catcom.2018.06.022"
}

@article{feng2018,
  title="Robust and Conductive Two-Dimensional Metal-Organic Frameworks with Exceptionally High Volumetric and Areal Capacitance",
  journal="Nat. Energy",
  author="D. Feng and T. Lei and M. R. Lukatskaya and J. Park and Z. Huang and M. Lee and L. Shaw and S. Chen and A. A. Yakovenko and A. Kulkarni and J. Xiao and K. Fredrickson and J. B. Tok and X. Zou and Y. Cui and Z. Bao",
  volume="3",
  pages="30--36",
  year="2018",
  doi="10.1038/s41560-017-0044-5"
}

@article{dong2024,
author={J. Dong and D. W. Boukhvalov and C. Lv and M. G. Humphrey and C. Zhang and Z. Huang},
title = {Enhancing Oxygen Evolution Reaction Performance of Metal-Organic Frameworks through Cathode Activation},
journal = {ChemSusChem},
volume = {17},
pages = {e202401176},
year = {2024},
doi = {10.1002/cssc.202401176}
}

@article{chan2025bi, 
title = {A Bimetallic {2D} {NiFe} {MOF/N-doped} Reduced Graphene Oxide as a Bifunctional Oxygen Catalyst for Rechargeable {Zinc-air} Batteries}, 
author = {Y.-P. Chan and C.-S. Huang and C.-Y. Lai and Y.-C. Chen and D. Chang and Y.-W. Chuang and K.-F. Tu and C.-N. Yeh}, 
journal = {Appl. Surf. Sci.}, 
volume = {692}, 
pages = {162720}, 
year = {2025}, 
doi = {10.1016/j.apsusc.2025.162720}, 
}

@article{fu2025fe, 
Title={{Fe/Co} Bimetal-Containing Carbon Prepared from a {2D} Metalloporphyrin-Based {MOF} for the Optimal {ORR/OER} Bifunction and Its Applicationin {Zn-air} Batteries},  
author={F. Zhen and Z. Hongyan and L. Xue and L. Wenjuan and S. Hao and S. Zhuang and F. Linlin and J. Tenglong and C. Wenmiao and C. Yanli}, 
journal = {ACS Appl. Energy Mater.}, 
volume = {8}, 
pages = {1051-1059}, 
year = {2025}, 
doi = {10.1021/acsaem.4c02571}, 
}

@article{wei2024un, 
title = {Unraveling the Intrinsic Mechanism of High-Performance Two-dimensional Conjugated Metal-Organic Frameworks for {ORR/OER} Through Theoretical Investigation}, 
author = {X. Wei and S. Cao and S. Cheng and C. Lu and X. Chen and X. Lu and X. Chen and F. Dai}, 
journal = {ACS Mater. Lett.}, 
volume = {6}, 
number = {8}, 
pages = {3496--3504}, 
year = {2024}, 
doi = {10.1021/acsmaterialslett.4c00961}, 
}

@article{das2023bi,  
title  = {Bifunctional Electrocatalytic Activity of Two-Dimensional Metallophthalocyanine-Based Metal-Organic-Frameworks for Overall Water Splitting: A {DFT} Study},  
 author = {P. Das and B. Ball and P. Sarkar},  
  journal = {ACS Catal.},  
  volume = {13},  
  number = {24},  
  pages  = {16307--16317},  
  year   = {2023},  
  doi    = {10.1021/acscatal.3c03967},  
}

@article{xu2023two,  
  title  = {Two-Dimensional Metal-Organic Frameworks as Bifunctional Electrocatalysts for the Oxygen Evolution Reaction and Oxygen Reduction Reaction ({OER/ORR}): A Theoretical Study},  
  author = {F. Xu and Z. Gao and Z. Ge and H. Ma and H. Ren and H. Zhu and Y. Chi and W. Guo and W. Zhao},  
  journal = {Phys. Chem. Chem. Phys.},  
  volume = {25},  
 number = {26},  
pages  = {17508--17514},  
year   = {2023},  
 doi    = {10.1039/D3CP01193C},   
}

@article{zhou2024me,  
 title  = {Metal and Ligand Modification Modulates the Electrocatalytic {HER}, {OER}, and {ORR} Activity of {2D} Conductive Metal-Organic Frameworks},  
  author = {Y. Zhou and L. Sheng and L. Chen and W. Zhao and W. Zhang and J. Yang},  
  journal = {Nano Res.},  
  volume = {17},  
   number = {9},  
   pages  = {7984--7990},  
   year   = {2024},  
   doi    = {10.1007/s12274-024-6813-0},  
 }

@article{wang2023spin,
author={Y. Wang and J. Jiang and J.-J. Zou and W. Mi},
title = {Spin-Gapless Semiconductors and Quantum Anomalous Hall Effects of Tetraazanaphthotetraphene-Based Two-Dimensional Transition-Metal Organic Frameworks on Spintronics and Electrocatalysts for {CO$_2$} Reduction},
journal = {ACS Appl. Electron. Mater.},
volume = {5},
pages = {1243--1251},
year = {2023},
doi = {10.1021/acsaelm.2c01694}
}

@article{hoyos-sinchi2025,
author={V. Hoyos-Sinchi and P. H. Souza and W. Orellana},
title = {Computational Insights into Two-Dimensional {M$_3$(HHTP)$_2$} Metal-Organic Frameworks as {ORR/OER} Electrocatalysts},
journal = {J. Phys. Chem. C},
volume = {129},
pages = {14002},
year = {2025},
doi = {10.1021/acs.jpcc.5c03531}
}

@article{youxiwang2025,
author={Y. Wang and L. Hua and Z. Li},
title = {Engineering Bifunctional Oxygen Electrocatalysts in {2D} Conjugated Metal-Organic Frameworks: Theoretical Perspectives on Metal-Ligand Synergistic Effects},
journal = {J. Phys. Chem. Lett.},
volume = {16},
pages = {11206},
year = {2025},
doi = {10.1021/acs.jpclett.5c02635}
}

@article{chen2021co,
author={K. Chen and C. A. Downes and E. Schneider and J. D. Goodpaster and S. C. Marinescu},
title = {Improving and Understanding the Hydrogen Evolving Activity of a Cobalt Dithiolene Metal-Organic Framework},
journal = {ACS Appl. Mater. Interfaces},
volume = {13},
pages = {16384},
year = {2021},
doi = {10.1021/acsami.1c01727}
}

@article{chen2021cofe,
author={K. Chen and C. A. Downes and J. D. Goodpaster and S. C. Marinescu},
title = {Hydrogen Evolving Activity of Dithiolene-Based Metal-Organic Frameworks with Mixed Cobalt and Iron Centers},
journal = {Inorg. Chem.},
volume = {60},
pages = {11923},
year = {2021},
doi = {10.1021/acs.inorgchem.1c00900}
}

@article{orellana2026,
author={W. Orellana},
title = {Ligand-dependent stability and {ORR/OER} activity of single-layer metal-organic frameworks},
journal = {Electrochim. Acta},
volume = {558},
pages = {148589},
year = {2026},
doi = {10.1016/j.electacta.2026.148589}
}

@article{shao2019,
author={Y. Shao and J.-P. Dodelet and G. Wu and P. Zelenay},
title = {{PGM}-free cathode catalysts for {PEM} fuel cells: A mini-review on stability challenges},
journal = {Adv. Mater.},
volume = {31},
pages = {1807615},
year = {2019},
doi = {10.1002/adma.201807615}
}

@article{menga2024,
author={D. Menga and Y.-S. Li and A. M. Damjanovic and O. Proux and F. E. Wagner and T.-P. Fellinger and H. A. Gasteiger and M. Piana},
title = {On the Stability of an Atomically-Dispersed {Fe--N--C} {ORR} Catalyst: An In Situ {XAS} Study in a {PEMFC}},
journal = {ChemElectroChem},
volume = {11},
pages = {e202400228},
year = {2024},
doi = {10.1002/celc.202400228}
}

@article{xie2025,
author={X. Xie and B. Li and P. Xu and M. T. Sougrati, R. Garcia-Serres and D. A. Cullen and others},
title = {Unravelling the Stability Stressors of Atomically Dispersed {Fe--N--C} Oxygen Reduction Catalysts},
journal = {J. Am. Chem. Soc.},
volume = {147},
pages = {48117},
year = {2025},
doi = {10.1021/jacs.5c15451}
}

@article{chen2025,
author={W. Chen and Y. Qin and C. Heng and S. Wang and M. Ge and Y. Li and X. Wan and X. Li and J. Shui and Y. Su and and D. Su},
title = {Degradation of {FeNC} Electrocatalysts for Acidic and Alkaline Oxygen Reduction},
journal = {J. Am. Chem. Soc.},
volume = {147},
pages = {35730},
year = {2025},
doi = {10.1021/jacs.5c11985}
}

@article{kobayashi2025,
author={R. Kobayashi and S. Honma and J. Ozaki},
title = {Quantitative detection of hydroxyl radical-mediated degradation in {Fe--N--C} catalysts during the oxygen reduction reaction},
journal = {Carbon Reports},
volume = {4},
pages = {270},
year = {2025},
doi = {10.7209/carbon.040406}
}

@article{xia2021,
author={D. Xia and C. Yu and Y. Zhao and Y. Wei and H. Wu and Y. Kang and J. Li and L. Gan and F. Kang},
title = {Degradation and regeneration of {Fe--N$_x$} active sites for the oxygen reduction reaction: the role of surface oxidation, 
{Fe} demetallation and local carbon microporosity},
journal = {Chem. Sci.},
volume = {12},
pages = {11576},
year = {2021},
doi = {10.1039/d1sc03754d}
}

@article{nali2023,
author={N. Li and L. Li and J. Xia and M. Arif and S. Zhou and F. Yin and G. He and H. Chen},
title = {Single-atom {Co-N$_4$} catalytic sites anchored on {N}-doped ordered mesoporous carbon for excellent {Zn}-air batteries},
journal = {J. Mater. Sci. Tech.},
volume = {139},
pages = {224},
year = {2023},
doi = {10.1016/j.jmst.2022.07.058}
}

@article{shantharaja2023,
author={Shantharaja and Giddaerappa and L. K. Sannegowda},
title = {Phthalocyanine based metal-organic framework with carbon nanoparticles as hybrid catalyst for oxygen reduction reaction},
journal = {Electrochim. Acta},
volume = {456},
pages = {142405},
year = {2023},
doi = {10.1016/j.electacta.2023.142405}
}

@article{kim2021,
author={Y. Kim and D. Kim and J. Lee and L. Y. S. Lee and D. K. P. Ng},
title = {Tuning the Electrochemical Properties of Polymeric Cobalt Phthalocyanines for Efficient Water Splitting},
journal = {Adv. Funct. Mater.},
volume = {31},
pages = {2103290},
year = {2021},
doi = {10.1002/adfm.202103290}
}

@article{yuanli2024,
author={Y. Li and M.-Y. Chen and B.-A. Lu and H.-R. Wu , J.-N. Zhang },
title = {Unravelling the role of hydrogen peroxide in {pH}-dependent {ORR} performance of {Mn-N-C} catalysts},
journal = {Appl. Catal. B-Environ.},
volume = {342},
pages = {123458},
year = {2024},
doi = {10.1016/j.apcatb.2023.123458}
}

@article{zhang2018,
author={M. Zhang and B.-H. Zheng and J. Xu and N. Pan and J. Yu and M. Chena and H. Cao},
title = {Fewer-layer conductive metal–organic nanosheets enable ultrahigh mass activity for the oxygen evolution reaction},
journal = {Chem. Commun.},
volume = {54},
pages = {13579},
year = {2018},
doi = {10.1039/c8cc08156e}
}
\end{document}